# Coaxial *p*- and *n*-type carbon nanotube transistors with dopant-selective coatings

H. B. Peng, J.A. Golovchenko*

*Division of Engineering and Applied Sciences, Harvard University, Cambridge, MA 02138*

Abstract

Carbon nanotube field-effect transistors (FETs) with passivated coaxial gate structures have been fabricated after growth of contacted suspended single wall nanotubes (SWNTs) and subsequent coating with gate dielectrics. Electron Fabry-Perot interferences are observed in the FETs indicating ballistic transport in the coated structures. Reliable production of high-performance SWNT FETs has been obtained by combining patterned growth of SWNT arrays with feed-back controlled electrical breakdown to select desired semiconducting SWNTs, making large-scale SWNT FET fabrication achievable. P-type and n-type passivated SWNT FETs have been realized through dopant-selective nanotube coatings, which enables the fabrication of active circuits based on complementary device structures.

*Also at: Department of Physics, Harvard University, Cambridge, MA 02138*

Carbon nanotubes have attracted intense attention for their potential application to future nanoelectronic technologies following the demonstration of nanotube field effect transistors (FETs). [1-6] Large scale production of FETs seems promising via patterned growth of single wall nanotubes (SWNTs) by CVD (chemical vapor deposition) using nanoparticles in solution [7] or low-coverage solid Fe [8] as catalysts. But it is still a major challenge to selectively grow semiconducting SWNTs (s-SWNTs) which are needed for FETs. Here we show that high-yield *p*- and *n*-type SWNT FETs with passivated coaxial gate structures can be fabricated after CVD growth of suspended SWNT arrays using solid Fe catalysts. Fabry-Perot interferences of electron waves are observed in the FETs at low temperatures indicating ballistic transport in the coated structures. In addition, we show that reliable production of high-performance SWNT FETs can be obtained by combining patterned growth of SWNT arrays with feed-back controlled electrical breakdown to select desired semiconducting nanotubes. Using these strategies, passivated p-type and n-type SWNT FETs are obtained with selective buffer layer coatings, which enables the fabrication of active circuits based on complementary device structures.

In this work, electrically contacted suspended SWNTs are CVD grown across a gap etched all the way through self-supporting silicon nitride membranes using solid Fe catalysts.[8,9] The suspended SWNTs are then coaxially coated with dielectric silicon nitride by low-pressure chemical vapor deposition (LPCVD) [9] to make *p*-type FET structures after subsequent optical lithography and lift-off processing to fashion metal gates electrodes that cover the nitride-coated SWNTs as well as parts of the coated source (drain) electrodes (Fig. 1a). (*n*-type devices will be discussed later.) The LPCVD process is crucial for creating a uniform gate dielectric around the suspended SWNTs [9].

We use SWNT growth parameters [9] so that most FETs initially have an array of several suspended SWNTs bridging the source and drain electrodes that include both semiconducting and metallic SWNTs. Fig. 1b shows the transfer characteristics (source-drain current $I_{DS}$ versus gate voltage Vg) for a device with multiple SWNTs ~2 μm in length and a dielectric silicon nitride coating 70 nm in radius. At $V_{DS}$=0.1V, the subthreshold swing is S ~ 500 mV per decade and the current ON-OFF ratio $I_{ON}/I_{OFF}$ is ~2000, which indicates semiconducting SWNTs dominate but there are still some contributions from metallic SWNTs (m-SWNTs) to the remaining current in the OFF state. We note that the mixed contribution of both semiconducting and metallic components is very common in our initial FET structures containing SWNT arrays.

The possibility of using electrical breakdown to modify multi-walled nanotubes and SWNT ropes was investigated previously. [4] We show here that controlled electrical breakdown is effective for eliminating metallic SWNTs in our FET structures, even though the nanotubes are encased in gate dielectric and even if the metallic components dominate the device characteristics initially. Fig. 2a and 2b show the characteristics of an initial FET structure with multiple SWNTs ~1μm in length and a silicon nitride coating of 100 nm radius. The metallic SWNTs dominate the electronic transport and result in a very low ON-OFF ratio of only ~3. The nitride-coated semiconducting SWNTs are p-type. Depletion of hole carriers occurs at positive gate voltage and inversion to n-type occurs at gate voltages $V_g$ greater than +1 V. The key to the electrical breakdown method is to protect semiconducting SWNTs from discharge damage by gating to avoid large current flows through them. Depleting the desired semiconducting SWNTs by a suitable $V_g$ results in large destructive current flow mainly through the metallic or most highly-conducting SWNTs under high source drain voltage. Fig. 2c shows such a breakdown process of SWNTs with Vg=+10 V and $V_{DS}$=15 V for the sample shown in Fig. 2a.

After this process, the OFF state current was reduced (Fig. 2d), and no inversion occurred at Vg = +3V, which shows the breakdown of some large diameter semiconducting SWNTs, as well as metallic SWNTs. When the temperature is lowered to 7 K, pronounced oscillations of source-drain current appear in the $I_{DS}$ versus gate voltage $V_g$ data (Fig. 2e). These can be explained as electron Fabry-Perot interferences [10, 6] inside the semiconducting SWNTs in its ON state. The inset of Fig. 2e diagrammatically illustrates how such an interference effect comes about for semiconducting nanotubes. The oscillations are significant in the ON state at negative gate voltage. They essentially disappear in the OFF state at positive gate voltage. Fig. 2f shows the linear conductance versus gate voltage near $V_g$ ~2V for different source-drain bias voltages. The interference patterns are well reproducible while scanning $V_g$ back and forth. As seen clearly, the conductance peaks shift to more negative $V_g$ as $V_{DS}$ is increased negatively. A possible explanation is that as $V_{DS}$ is increased, there are more and more states with different wave vectors *k* contributing to the interference. Consequently, the peak position shift in concert with the shift of the average contribution of all these states. Also, the conductance oscillations smear out as $V_{DS}$ increases. Note that Fabry-Perot interference in semiconducting SWNTs was reported only recently in nearly Schottkey barrier free Pd-contacted SWNTs. [6] Our observation of the Fabry-Perot type interference in passivated s-SWNTs suggests that the suspended SWNTs are high quality with few intrinsic defects.

We note that during the process to optimize FET performance, the electrical breakdown of SWNTs usually happens rapidly after application of a high $V_{DS}$, and it is easy to damage all SWNTs. However, high-performance FET devices can be reliably obtained using an automatic feedback control system that monitors the source-drain current during the "burning" process. If $I_{DS}$ drops to a certain threshold value, the feedback system resets the source-drain bias voltage to zero so that the remaining SWNTs will be protected. The use of an automatic feedback control is crucial to produce FET devices reliably from as-grown SWNT arrays. At high source-drain bias, optical phonon scattering has been suggested to play an important role in the electron transport of both m-SWNTs and s-SWNTs. [11,12] Typically, the current-carrying capability of a single SWNT is tens of μA. Therefore, to obtain at least one SWNT protected for a FET device, we usually choose threshold currents ~10 μA for triggering the automatic feedback control.

Moreover, by choosing different gate voltage biasing, it is possible to select s-SWNTs with desired band gaps for optimizing the FET performances, since the as-grown SWNTs often have a distribution of diameters $d$, and thus a distribution of band-gaps $Eg \propto 1/d$. The strategy is to carry out an initial electrical breakdown by biasing the FETs at relatively high positive gate voltage to destroy the large-diameter (small band gap) s-SWNTs showing inversion and metallic SWNTs. This is followed by stimulating further electrical breakdown by biasing at selected lower gate voltages which still leaves the desired nanotubes in the OFF state. The second process gets rid of small-diameter s-SWNTs which still have some conduction at the desired OFF-state gate voltage. With this two step strategy, the ON-OFF ratio could be controllably improved. Fig. 3a shows one sample produced in this way. We first biased the gate at Vg=+10 V to do the initial electrical breakdown processes, followed by a final breakdown process biasing at Vg=+3V and $V_{DS}$=-30V with a threshold current 20 µA for the automatic feedback control. As seen from Fig. 3a, the ON-OFF ratio is greatly improved, from only ~ 3 for the initial FET to more than $10^5$, and the subthreshold swing is S ~ 150 mV per decade for the final device.

As shown above, CVD growth of SWNT arrays via solid Fe catalyst combined with feedback controlled electrical breakdown is effective for obtaining high-performance FETs. We emphasize that the growth yield for suspended SWNTs crossing gaps made through the membrane is high [8]. For a gap < 3µm in width, we have nearly 100% yield of SWNT devices with SWNT arrays bridging the electrodes. Therefore, by combining the post-growth electrical breakdown method, large-scale production of SWNT FETs is achievable, without addressing the difficult problem of selectively growing only semiconducting SWNTs.

Reliable n-type coaxial SWNT FETs can also be obtained by using Cr doping and an aluminum oxide coating as a buffer layer on the nanotube. It is well known that as-grown s-SWNTs are p-type due to the exposure to ambient air. Since the ability to produce both p-type and n-type SWNT FETs is important for constructing complementary electronic circuit, attempts have been made to obtain n-type SWNTs.[13-15] We show here that dopant-selective coating of suspended SWNTs offers a reliable way to produce stable n-type passivated coaxial SWNT FETs.

Fig. 3b shows the transfer characteristics of a coaxial FET device, fabricated by coating SWNTs with a layer of 1 nm Cr (by thermal evaporation) and 20 nm $Al_2O_3$ (by atomic layer deposition), followed by a passivating coating of 70 nm silicon nitride by LPCVD as the main gate dielectrics. The Cr metal was thermally evaporated to coat the bare SWNTs from the back no-electrode side of the membrane, to avoid shorting the source-drain electrodes. As seen clearly, the FET is n-type with higher conductance at positive gate voltage. The FET characteristics (Fig. 3b) were measured after destroying metallic SWNTs by the electrical breakdown method. The behavior of the initial FET indicated dominant contributions from metallic components. It is worthwhile to mention that although LPCVD serves well to directly coat the suspended bare SWNTs uniformly, other processes such as atomic layer deposition (ALD) [16,17] are not able to uniformly coat suspended SWNTs directly. Fig. 3c shows the coating behavior of $Al_2O_3$ on bare SWNTs by the ALD process. Most parts of the bare SWNTs are not coated, but localized nucleated spheres of $Al_2O_3$ are clearly visible. The $Al_2O_3$ probably nucleates at sites on SWNTs where there are defects (such as kinks) or contaminations. This is also consistent with a weak interaction between the SWNT surface and the precursors, especially $OH^-$ groups, that initiate the ALD process. By precoating SWNTs with a thin layer of Cr which also plays a role in doping, it is then possible to coaxially coat SWNT uniformly, as seen in Fig. 3d. This general approach enables the fabrication of passivated coaxial structures based on SWNTs.

The combination of Cr and the $Al_2O_3$ layer together plays a crucial role in obtaining the n-type FET. If we only use 1 nm Cr coating and LPCVD silicon nitride as the gate dielectrics without the $Al_2O_3$ layer, the FET behavior is still p-type, despite the fact that after Cr coating alone the suspended SWNTs usually have lower conductance (presumably from the depletion of hole carriers by the electron doping by Cr). (Thereafter the conductance of SWNTs coated with ~ 1 nm Cr increases gradually again when the samples are exposed to air, which shows its sensitivity to the surrounding environment.) The $Al_2O_3$ layer serves as a barrier to screen the influence of silicon nitride from the Cr doping layer. For device consideration, a thin Cr coating and a barrier layer to preserve the n-type behavior allows more choices of efficient (e.g., high-*k*) gating dielectrics. Further investigations with combinations of different gating materials and dopants will certainly open a new window to functionalize the SWNT circuits and optimize device performances.

We thank Erli Chen for experimental assistance. This work was supported by D.O.E. and the N.S.F.

# Figure Captions

Fig. 1. (a) Schematic cross-sectional view of the SWNT FET devices. (b) Transfer characteristics of a coaxial FET device as built from suspended SWNTs ~ 2µm in length with ~70 nm radial silicon nitride gate dielectrics. Inset: SEM image of the device after silicon nitride coating, but before the patterning of the gate metal. Scale bar: 1 µm.

Fig. 2. (a) Transfer characteristics of a coaxial FET device as built from suspended SWNTs ~ 1µm in length. The gate dielectrics is LPCVD silicon nitride ~ 100 nm in radius. (b) $I_{DS}$-$V_{DS}$ versus Vg at room temperature for the same device. (c) $I_{DS}$ versus time during an electrical breakdown process in vacuum under $V_{DS}$=15V and Vg=+10V. (d) $I_{DS}$ versus $V_g$ under $V_{DS}$=10 mV at room temperature for the device before and after electrical breakdown. (e) $I_{DS}$ versus $V_g$ under $V_{DS}$=14 mV at T = 7 K for the device in d after electrical breakdown. Inset: schematic diagram of the energy dispersion relation of a semiconducting SWNT. Interferences associated with electronic states at $k_1$ and $k_2$ modulate the electrical conduction and are responsible for the Fabry-Perot interference. (f) Linear source-drain conductance G=$I_{DS}$/$V_{DS}$ versus $V_g$ under various $V_{DS}$ at T=7K for the device in e.

Fig. 3. (a) Room temperature transfer characteristics of a FET device built from coating suspended SWNTs ~ 1µm in length with 100 nm silicon nitride in radius as gate dielectrics. The data was recorded in vacuum after multiple electrical breakdown

processes with automatic feedback control. The initial breakdown processes were carried out under a gate biasing Vg=+10V. And the final breakdown was under Vg=+3V and $V_{DS}$= -30V with a threshold $I_{DS}$=20 μA for triggering the automatic feedback control. Inset: $I_{DS}$ versus Vg for the original device before electrical breakdown. (b) Room temperature transfer characteristics of a n-type FET after tuned by the electrical breakdown method. The SWNTs in the device are ~2μm in length. A 70 nm LPCVD silicon nitride gate dielectrics surrounds a combined buffer layer of $Al_2O_3$ (20nm) / Cr (1nm). The $Al_2O_3$ was deposited by ALD at 250 ºC with precursors $Al(CH_3)_3$ and $H_2O$. Cr coating of the bare SWNTs was by thermal evaporation. (c) TEM image of as-grown suspended SWNTs after an ALD process which yields 15nm $Al_2O_3$ on planar Si substrate. Scale bar: 50 nm. (d) TEM image of $Al_2O_3$-coated suspended SWNTs by ALD after a precoating of 4nm thermally evaporated Cr. Scale bar: 50 nm. Inset: TEM image of a nanotube coated with 4nm Cr.

Peng Fig. 1

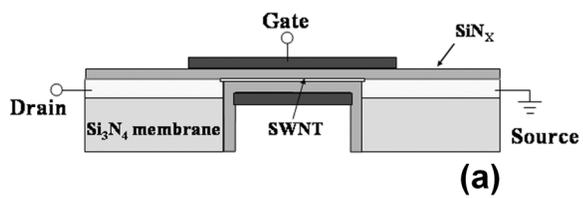

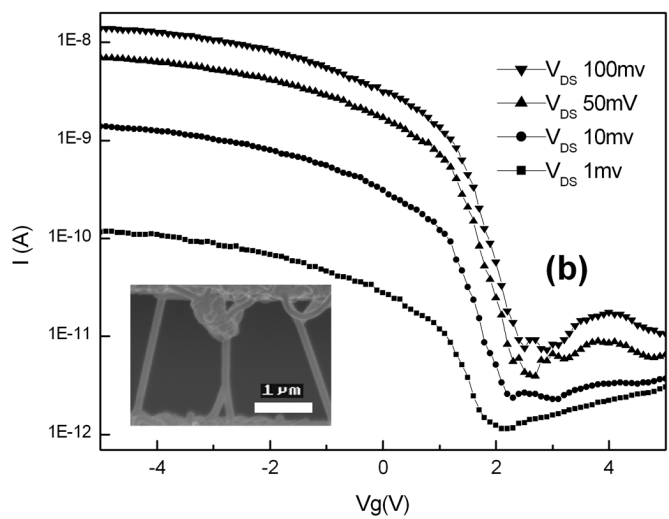

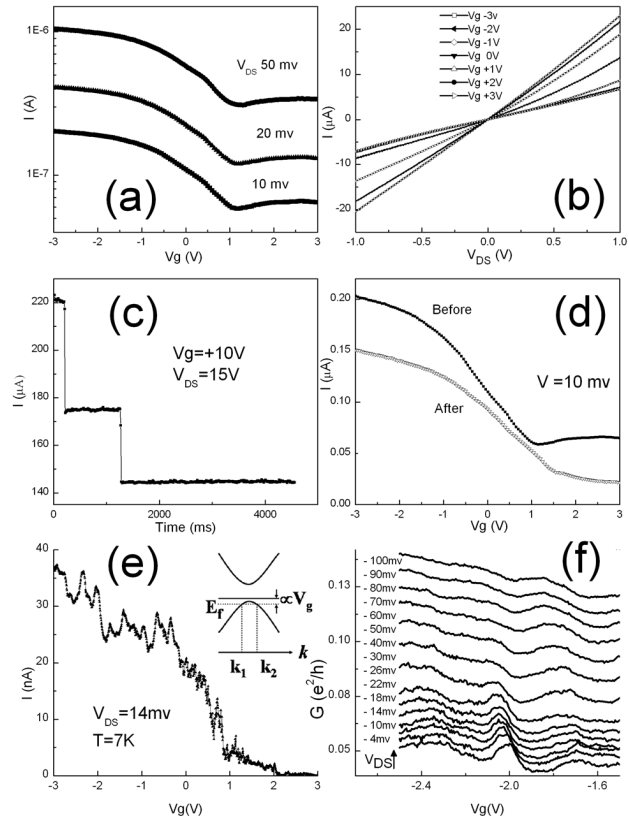

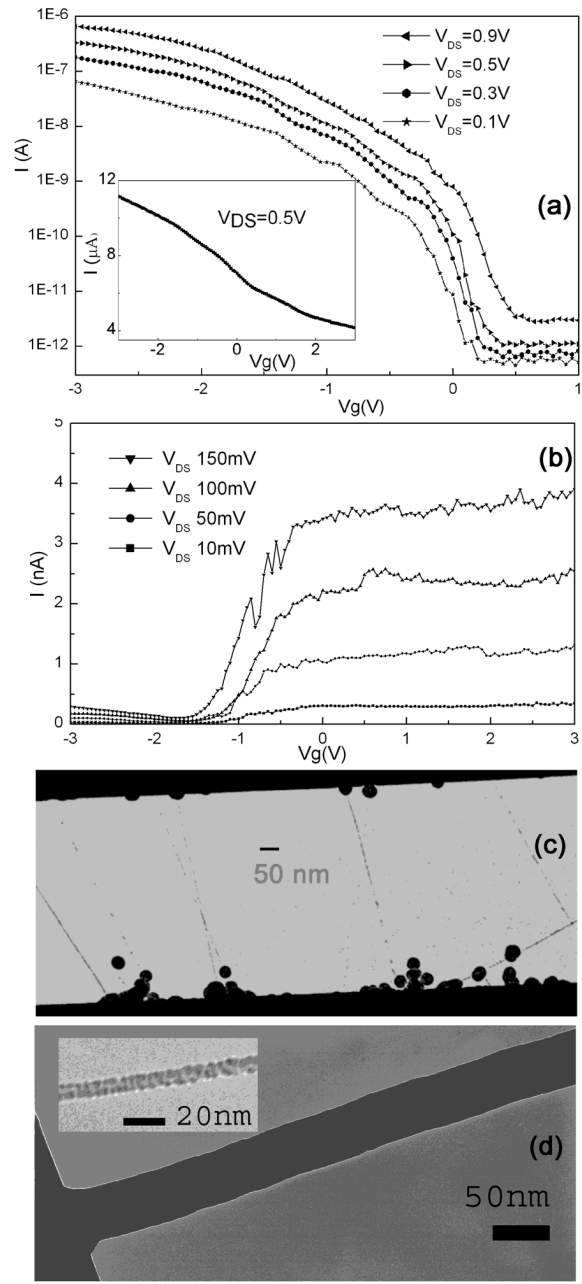